\documentstyle[12pt,psfig,twoside]{article} %use this if also
\begin{document}
\hbadness=10000
\pagenumbering {arabic}
\pagestyle{myheadings}
\markboth{J. Letessier, A. Tounsi, U. Heinz, J. Sollfrank and 
          J. Rafelski}
         {High entropy phase}
%%%%%%%%%%%%%%%%%%%%%%%%%%%%%%%%%%%%%%%%%%%%%%%%%%%%%%%%%%%%%%%%%%%%%%
\title{\bf EVIDENCE FOR A HIGH ENTROPY PHASE IN NUCLEAR COLLISIONS}
%%%%%%%%%%%%%%%%%%%%%%%%%%%%%%%%%%%%%%%%%%%%%%%%%%%%%%%%%%%%%%%%%%%%%%
\author{J. Letessier$^{(1)}$, A. Tounsi$^{(1)}$, 
U. Heinz$^{(2)}$\thanks{Work supported by DFG, BMFT, GSI, 
and NATO, grant CRG 910991} , J. Sollfrank$^{(2)}$, 
and J. Rafelski$^{(3)}$\thanks{Work supported by DOE, grant DE-FG02-92ER
40733, and 
NATO, grant CRG 910991}
\\
\footnotesize\it
$^{(1)}$Laboratoire de Physique Th\'eorique et Hautes 
Energies\thanks{ Unit\'e associ\'ee au CNRS} , Universit\'e Paris 7
\\ \footnotesize\it
Tour 24, 5\`e \'et., 2 Place Jussieu, F-75251 Paris CEDEX 05, France\\
\footnotesize\it
$^{(2)}$Institut f\"ur Theoretische Physik, Universit\"at Regensburg,
D-8400 Regensburg, Germany\\
\footnotesize\it
$^{(3)}$Department of Physics, University of Arizona, Tucson, AZ 85721\\
\\(October 1992)}
 
\date{} %Deleting this command produces today's date.
\maketitle

%%%%%%%%%%%%%%%%%%%%%%%%%%%%%%%%%%%%%%%%%%%%%%%%%%%%%%%%%%%%%%%%%%%%%%

\begin{abstract}
\noindent
We determine the entropy per baryon content of the central reaction
region in terms of the charged particle multiplicity. 
We study the consistency of
our findings with recent data on strange anti-baryon production at 200
GeV\,A in S$\to$A collisions (A$\sim$200) assuming formation of a central
fireball. Hadron gas models which 
do not invoke strong medium modifications of hadron masses do not provide
enough entropy and are inconsistent with the
combined experimental results. In contrast the quark-gluon plasma
hypothesis explains them naturally.

\end{abstract}

\vspace*{0.5cm}

\begin{center}
Published in {\it Phys. Rev. Letters} 70 (1993) 3530--3533.
\end{center}
%%%%%%%%%%%%%%%%%%%%%%%%%%%%%%%%%%%%%%%%%%%%%%%%%%%%%%%%%%%%%%%%%%%%%%
\vfill
{\bf PAR/LPTHE/92--37}\hfill\\
{\bf TPR-92-39}\hfill\\
{\bf AZPH-TH/92-30}
         \hfill{\bf September 30, 1992}
\eject

%%%%%%%%%%%%%%%%%%%%%%%%%%%%%%%%%%%%%%%%%%%%%%%%%%%%%%%%%%%%%%%%%%%%%%
 
One of the fundamental differences between the two phases of dense
hadronic matter often referred to as hadronic gas (HG) and quark-gluon
plasma (QGP) is the entropy content per baryon which can be observed in
the final particle multiplicity. At fixed temperature $T$ (as, for
example, given by the observed slope of the momentum spectra of the
emitted particles) the QGP is the phase with the higher entropy; this
difference occurs because of the liberation of the color degrees of
freedom in the color deconfined QGP phase. In the domain of temperatures
and flavor chemical potentials explored in the recent  experiments with
200 GeV\,A sulphur-ion collisions this difference is  larger than a
factor two \cite{LTR92}.

Here we note (for more details see \cite{LTHSR}) that it is possible to
relate the observed strange particle data with multiplicity data and
obtain a specific  
measurement of the entropy content of the central
rapidity source of strange baryons and anti-baryons. We show that this
leads to a clear incompatibility of the hadronic gas reaction model with
the observations which require a considerably larger specific entropy in
the source. On the other hand we observe that the required entropy
content is in agreement with the QGP hypothesis. This assertion
supplements the earlier finding \cite{Raf91} that the strange particle
source has, in spite of its sizeable baryon chemical potential, a
symmetric strangeness phase space (as is generally true for a QGP), which
is at the same time nearly fully saturated \cite{fn1},  
properties difficult to understand  in normal hadronic matter.

We note that in 200 GeV\,A sulphur reactions involving heavy nuclear
targets the observed particle spectra, after correcting for resonance
decay effects \cite{sollfrank}, resemble thermal distributions possibly
including a collective flow component \cite{SH92}. An interpretation in
terms of a generalized thermal model which assumes local thermal
equilibrium of all components, chemical equilibrium of the gluon and
light quark abundances, and allows for incomplete chemical equilibrium
for the strange quark abundance, appears to be consistent with the
observed particle ratios \cite{Raf91,LTHSR}. 
 
The central fireball from which the particles emerge is described by its
temperature $T$ and by the chemical potentials $\mu_i$ of the 
different conserved quark flavors $u,d,s$ (or by the corresponding 
fugacities $\lambda_i=\exp(\mu_i/T)$). %Since the conserved quantum 
%numbers of hadrons are simply the sum of the corresponding quantum 
%numbers of their quark constituents, 
The hadronic chemical potentials 
are given as the sum of constituent quark chemical potentials. The 
strong and electromagnetic interactions do not mix quark flavors, and 
the net numbers of $u,d,$ and $s$ quarks are separately conserved on 
the time scale of hadronic collisions. For the strange flavor we also 
introduce the phase space saturation factor $\gamma_{\rm s}$ 
\cite{Raf91,LTR92}. In that framework an analysis of the strange 
particle momentum spectra within the fireball model \cite{RRD92} found 
consistency of all strange particle measurements in S$\to$A collisions 
(with $A\sim200$) at 200 GeV\,A with an ``apparent temperature" 
(inverse slope parameter) of $T_{\rm apparent} = 210 \pm 10$ MeV. This 
value comprises in principle the effects of thermal motion and 
collective expansion flow, and it is still a matter of debate 
\cite{sollfrank,SH92} how these two effects superimpose each other in 
the heavy-ion data.  
 
Two extreme and sometimes complementary pictures of the dynamics can 
be imagined. In the first one assumes thermal evaporation or sudden 
disintegration without collective flow, such that the inverse slope 
reflects the true temperature, $T\simeq 210$ MeV. In the second 
approach one assumes $T=150$ MeV combined with a transverse flow 
velocity $\beta_{\rm f}=0.32$ at freeze-out, leading to the same 
apparent temperature via the blueshift effect. It has already been 
pointed out \cite{LTR92} that only the former scenario ($T>200$ MeV) 
can be consistent with a conventional {\it hadronic gas} constrained 
by the principle of strangeness conservation and the observed 
strange--anti-strange phase space symmetry $\mu_{\rm s} \simeq 0$,
unless a strong modification of the hadron gas equation of state 
is performed (see also \cite{Cley92,LTHSR}). 
%To shift the point of consistency between 
%the strangeness neutrality condition and the observed value $\mu_{\rm 
%s}=0$ towards lower temperatures would require massive modifications 
%of the hadron gas equation of state, e.g. in the form of medium 
%modifications of hadron masses. 
In the conventional HG picture, at a 
temperature $T=150$ MeV the strangeness constraint at $\mu_{\rm s}=0$ 
requires a very large value of the baryon chemical potential, 
$\mu_{\rm B}\sim 900$ MeV. This value is entirely incompatible with 
the experimental yield of strange anti-baryons as it implies for 
example a ratio $\overline{\Lambda} / \Lambda\simeq \exp[-{4\over3} 
\mu_{\rm B}/T] \simeq 0.0003$. This should be compared to the reported 
\cite{WA85} central rapidity result for S--W collisions triggered on 
high multiplicity which, after correction for the contamination from 
cascade decays, is $\overline{\Lambda} / \Lambda = 0.13 \pm 0.03$.  
Similarly, the ratio $\overline{\Xi}/\Xi\simeq \exp[-{2\over3}\mu_{\rm 
B}/T]$ would yield $\simeq 0.02$ instead of $\sim 0.4$ reported. Thus, 
if confirmed, the observation of $\mu_{\rm s}=0$ excludes the flow 
interpretation of the observed spectra within a conventional 
equilibrium HG picture \cite{SH92}, and within such a picture the 
measured slope of $\simeq 210$ MeV must be taken as a serious 
indicator of the true temperature of the source. However, at $T > 200$ 
MeV the notion of a hadron resonance gas develops numerous internal 
inconsistencies which alone raise doubts about the validity of such a 
hadron gas interpretation. In the context of a QGP true temperatures 
of 200--220 MeV are, of course, entirely possible, but in this case 
the observation $\mu_{\rm s} \simeq 0$ does not force us to accept the 
apparent slope directly as a temperature, since this value of 
$\mu_{\rm s}$ is assumed by a strangeness neutral QGP {\it at any 
temperature}. 
 
As a further argument against the interpretation of the data in terms 
of a HG state at $T \simeq 210$ MeV we show here that the 
corresponding entropy content of such a state is inconsistent with the 
available data on charged particle multiplicity, while no such problem exists 
if one instead assumes the formation of a high entropy quark-gluon 
phase. A first indication for a grave inconsistency in the HG 
interpretation is contained in the results discussed in 
Ref.\cite{Dav91}; a combined chemical analysis of the NA35 \cite{NA35} 
and WA85 \cite{WA85} data led these authors to assume an equilibrated 
HG at $T=170$ MeV and $\mu_{\rm B}=257$ MeV as the source for the 
emitted strange particles. With these parameters the observed $\pi/K$ 
ratio is underpredicted by nearly a factor 2, indicating an 
overabundance of pions and thus excess entropy in the data which are 
not compatible with the thermal HG model. 

To quantify this argument and allow for a phenomenological discussion 
of the entropy balance, we introduce the following easily measurable 
quantity: 
 \begin{equation} 
  D_{\rm Q}\equiv{N^+ - N^-\over N^+ + N^-}\, .
  \label{DQ} 
 \end{equation} 
on which we will base. 
$D_{\rm Q}$ is the ratio of net charge multiplicity to the total 
charged multiplicity. It can be determined (without identifying 
particles) with any tracking device within a magnetic field. As long 
as the experimental acceptance and efficiency is similar for particles 
with both signs of the electric charge, this ratio is insensitive to 
acceptance corrections. A simple and well-known estimate of this ratio 
in which only pions and nucleons are counted, assuming $N_{\pi^+} = 
N_{\pi^-} = N_{\pi^0} = N_\pi/3$, is given by 
 \begin{equation} 
  D_{\rm Q}^{\rm nonstrange} \approx {{\cal B}\over {N_\pi}}\ 
             {0.75 \over {1+0.75\, ({\cal B}/N_\pi)}} \, , \label{DQBP}
 \end{equation} 
where $\cal B$ is the number of baryons in the fireball. As we see 
$D_{\rm Q}$ is a measure of the baryon to pion ratio. We note that the 
estimate Eq.\,(\ref{DQBP}) is wrong by two partially compensating 
factors of order 2: both higher mass non-strange resonances as well as 
charged strange particles must be included at their observed level, 
and therefore we will study $D_{\rm Q}$ and its relation to the 
entropy content of the source numerically.  
 
\begin{figure}[t]\vspace{-1.8cm}
\centerline{\hspace{0.2cm}\psfig{figure=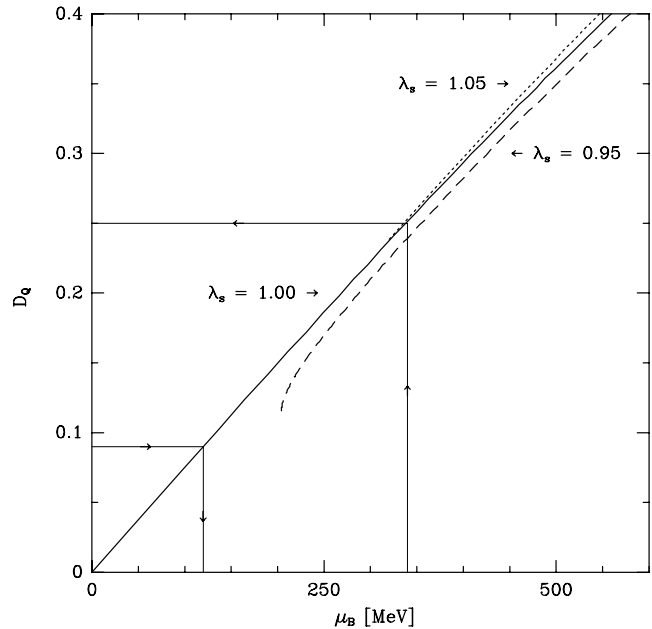  ,height=12.5cm}}
\vspace{-2cm}\caption{
$D_{\rm Q}$ as function of $\mu_{\rm B}$ for fixed $\lambda_{\rm
s}=1\pm0.05$ in a HG fireball with zero conserved strangeness.
\protect\label{F6}
 }
\end{figure}

We constrain in our calculations \cite{LTHSR} the domain of thermodynamic
parameters such that the net strangeness of the gas is $\langle s \rangle
- \langle \bar s \rangle \simeq 0$. Imposing the phase space symmetry
between strange and anti-strange particles by setting the strange quark
fugacity $\lambda_{\rm s}=1\pm0.05$, as required by the observed strange
baryon and anti-baryon ratios \cite{Raf91,LTHSR} we can eliminate the
temperature as a free parameter. We thus obtain a relation between the
computable quantity $D_{\rm Q}$ and the baryon chemical potential (baryon
density) of the hadronic gas which we show in Fig.\,\ref{F6}. Up to
$\mu_{\rm B} \leq 600$ MeV we observe a nearly linear behavior $D_{\rm Q}
\approx \mu_{\rm B} /1.3$ GeV. The vertical line in Fig.\,\ref{F6}
indicates the intercept with $\mu_{\rm B}=340$ MeV, the preferred value
obtained in the analysis of strange baryon multiplicities
\cite{Raf91,LTHSR}. $D_{\rm Q}$ for the hadronic gas, as indicated by the
horizontal line, is thus of the order 0.25.
%%%%%%%%%%%%%%%%%%%%%%%%%%%%%%%%%%%%%%%%%%%%%%%%%%%%%%%%%%%%%%%%%%%%%%
%%%%%%%%%%%%%%%%%%%%%%%%%%%%%%%%%%%%%%%%%%%%%%%%%%%%%%%%%%%%%%%%%%%%%% 
Experiment EMU05 \cite{EMU05} has determined the ratio (\ref{DQ}) as a
function of pseudorapidity in a sample of 15 high multiplicity ($>300$)
events of 200 GeV\,A S--Pb collisions which were obtained by placing a 
200 $\mu$ Pb foil in front of the emulsion track detector. In the 
central (pseudo)rapidity region they measure $D_{\rm Q}(y\simeq 
2.5\pm0.5) = 0.088 \pm 0.007$, which differs considerably from the HG 
based theoretical expectation above. Similar values for $D_{\rm Q}$ 
can be extracted from the results reported by the NA5 collaboration 
\cite{NA582} for p--p and p--$A$ reactions (for a comparison with 
$A$--$A$ collisions see also \cite{NA3591}); however, in this case the 
projectile contains no neutrons (which do not contribute to the 
numerator in Eq.~(\ref{DQ}) but produce particles which contribute to 
the denominator). A similar value for $D_{\rm Q}$ in p--p and $A$--$A$ 
collisions thus is an indicator for additional stopping of baryons at 
central rapidity. It is also worth pointing out that, although the 
global features represented by $D_{\rm Q}$ appear at first sight to be 
rather similar in p--p, p--$A$ and $A$--$A$ collisions, there are 
striking differences in certain particle ratios, in particular of 
strange particles, which show that $D_{\rm Q}$ alone is not sufficient 
to characterize the collision mechanisms.  

From Fig.\,\ref{F6} we see that within the thermal HG model the EMU05 
value of $D_{\rm Q}$ requires a baryon chemical potential of $\mu_{\rm 
B} \simeq 120$ MeV. Note that for $\lambda_{\rm s}\simeq 1$ this value 
of $\mu_{\rm B}$ translates into $\overline{\Lambda}/\Lambda \simeq 
\exp[-{4\over3} \mu_{\rm B}/T] \simeq 0.47$ instead of the measured 
value $0.13\pm 0.03$ for $m_\bot > 1.7$ GeV \cite{WA85}. We thus 
conclude that the hypothesis of a hadronic gas fireball for 200 GeV~A 
S--W collisions is glaringly inconsistent with the combined EMU05 and 
WA85 data on $D_{\rm Q}$ and the $\overline{\Lambda}/\Lambda$ 
ratio.  
 
We will now relate the observable $D_{\rm Q}$ to the entropy content 
of the fireball. If we assume that hadronization of some primordial 
hadronic phase results in hadrons with thermal momentum distributions 
(as the data seem to indicate), i.e. that at most chemical 
equilibrium, but not thermal equilibrium is broken by the 
hadronization process, then we are allowed to relate the specific 
entropy per baryon $\cal S/B$ of the primordial phase with the value 
$D_{\rm Q}$ for the resulting hadrons using the hadronic gas model for 
the fireball. Note that $\cal S/B$ can only increase during evolution 
of the primordial phase into hadrons; it is a nearly conserved 
quantity in all existing hadronization models (see, for example, 
Ref.~\cite{FRI83}). It is easy to check \cite{LTHSR} that the relation 
${\cal S/B}(D_{\rm Q})$ is nearly independent of the thermodynamic 
state of the hadron emitting source: the product of $D_{\rm Q}$ with 
$\cal S/B$ is essentially proportional to the entropy per emitted 
pion, cf. Eq.\,(\ref{DQBP}). Therefore $D_{\rm Q}$ is a direct 
measure of the specific entropy of the system.  
 
%%%%%%%%%%%%%%%%%%%%%%%%%%%%%%%%%%%%%%%%%%%%%%%%%%%%%%%%%%%%%%%%%%%%%%%%%%
\begin{figure}[t]\vspace{-1.8cm}
\centerline{\hspace{0.2cm}\psfig{figure=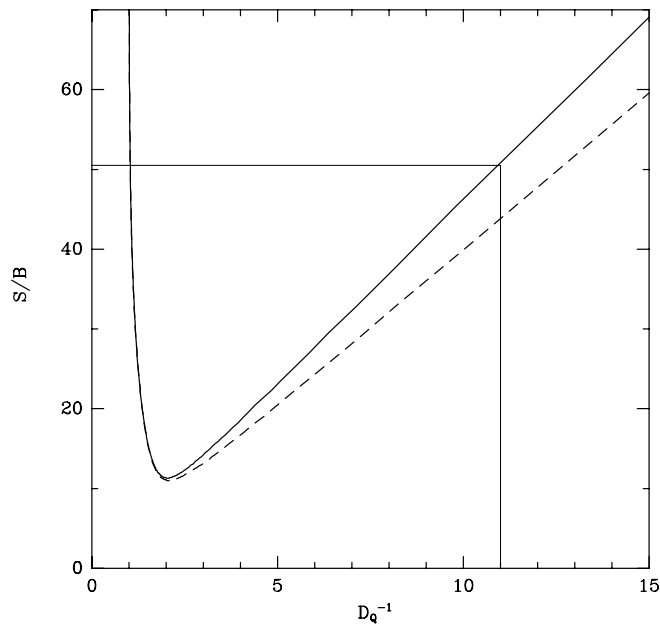  ,height=12.5cm}}
\vspace{-2cm}\caption{
Entropy per baryon ${\cal S/B}$ as a function of $D_{\rm Q}^{-1}$ for
fixed $\lambda_{\rm s}=1$ and conserved zero strangeness. Dashed: same
without strange particles.
\protect\label{F7bis}
 }
\end{figure}
%%%%%%%%%%%%%%%%%%%%%%%%%%%%%%%%%%%%%%%%%%%%%%%%%%%%%%%%%%%%%%%%%%%%%%%%%%
In Fig.\,\ref{F7bis} we show ${\cal S/B}(D_{\rm Q})$ computed with the 
observed value $\lambda_{\rm s}=1$ and for vanishing net strangeness.  
(The long-dashed line in this Figure is the result {\em without} 
strange particles, showing that the strange particle corrections are 
large: at fixed $D_{\rm Q}$, almost half of the entropy is contained 
in strange particles!). We see in particular that the measured value 
$D_{\rm Q} \simeq 0.09$ implies a specific entropy ${\cal S/B}\simeq 
50$. This has to be contrasted to a value of about 21 units per baryon 
which we would expect from the hadron gas value $D_{\rm Q}=0.25$ (see 
Fig.\,\ref{F6}). Thus the entropy extracted from the multiplicity 
measurements is by more than a factor 2 larger than the value provided 
by the hadron gas interpretation of the particle ratios! We can 
further strengthen our arguments for a high entropy phase by 
considering the temperature dependence of the relationship between 
$\cal S/B$ and the observable $D_{\rm Q}$. In order to reintroduce $T$ 
as a parameter we relax the requirement $\lambda_{\rm s}=1$ ($\mu_{\rm 
s}=0$) but continue in this calculation to request zero strangeness 
content of the fireball. For the entire range of temperatures between 
150 and 300 MeV the value $D_{\rm Q} \simeq 0.09$ implies ${\cal 
S/B}=50\pm4$. We verified that this observation remains in essence 
valid even if also the strangeness conservation requirement is 
slightly relaxed, as may be required if the fireball emits strange 
flavor asymmetrically. A complete description of these results will be 
presented in Ref.\,\cite{LTHSR}.  
 
Having shown that the measured value of $D_{\rm Q}$ points to the 
formation of a high entropy state, we proceed to further discuss its 
interpretation in terms of a QGP fireball. We use the perturbative 
equation of state where the presence of $\alpha_{\rm s}$ reduces the 
light quark and gluon effective degrees of freedom according to the 
first order perturbative formula. Along with most other work we leave 
the contributions of the massive strange quarks to entropy, energy, 
etc.  unaffected by these perturbative QCD interactions. In the QGP 
phase a fireball with zero strangeness always possesses $\mu_{\rm 
s}=0$. With $\mu_{\rm d}/T= 0.54$, $\mu_{\rm u}/T=0.51$ (allowing for 
the asymmetry between the $u$ and $d$ quarks, see \cite{LTHSR}) we 
find for $T=210$ MeV and a value of $\alpha_{\rm s}=0.6$ that the QGP 
energy density is 2.2 GeV/fm$^3$, the baryon density 0.27/fm$^3$ 
(nearly twice normal nuclear density), and the fully saturated strange 
quark density would be 0.64/fm$^3$. A non-negligible portion of the 
energy is contained in the strange quark pair density. A further small 
component is in the latent energy density of the vacuum which we took 
here to be $(170\ {\rm MeV})^4$.
While all these quantities depend rather 
strongly on the actual value of $T$, the entropy per baryon depends 
only on the ratio $\mu_{\rm B}/T$ (up to small corrections from the 
strange quark mass). Thus it is determined by the $\overline{\Lambda} 
/ \Lambda$ and $\overline{\Xi}/\Xi$ data \cite{WA85}, independent of 
how the $m_\bot$-spectra are interpreted in terms of temperature and 
flow, including the influence of resonance decays \cite{LTHSR,fn1}. We 
obtain ${\cal S}^{\rm QGP}/{\cal B}=47$ for $T/\mu_{\rm B} = 210/340 = 
0.62$. Changing $\alpha_{\rm s}$ from 0.6 to 0.4 changes ${\cal 
S}^{\rm QGP}/{\cal B}$ to 51, for $\alpha_{\rm s}=0.8$ we find ${\cal 
S}^{\rm QGP}/{\cal B} = 41$. A further uncertainty in the value of 
${\cal S}^{\rm QGP}/{\cal B}$ of the order of 10\% arises from the 
dependence of the strange quark component of the entropy on the ratio 
$m_{\rm s}/T$.  
 
Although the thermal picture becomes less reliable for smaller collision
systems, let us shortly consider a similar analysis for the results
obtained by the NA35 collaboration \cite{NA35,Wenig} in S--S interactions
at 200 GeV\,A. In this case we can base our discussion  only on the
reported charged particle rapidity densities \cite{Wenig}  and the
measured $\overline{\Lambda} / \Lambda$ ratio \cite{NA35}. S--S
interactions show at central rapidity visibly less baryon number stopping
than S--W collisions; from the data on $d(N^+-N^-)/dy$ and  $dN^-/dy$ in
\cite{Wenig} we read off a central rapidity value of  $D_{\rm Q}^{\rm
S-S}(y=3)=0.065$ with a probable error of order 25\%.  According to
Fig.\,\ref{F6} this implies a specific entropy $\cal S/B$ = 65--75. The
large magnitude of this number is due to the considerable degree of
transparency in this small collision system, resulting in a low central
baryon density. The raw $\overline{\Lambda}  / \Lambda$ ratio
(uncorrected for $\Xi$-decays) is strongly peaked at central rapidity and
rises there to a value $\overline{\Lambda} / \Lambda = 0.34$ (with an
error of order 50\%) \cite{NA35}. This implies a non-zero combination of
chemical potentials, $\mu_{\rm B}+1.5\mu_{\rm s} = (0.8 \pm 0.4)\,T$.
Unfortunately, the large error in this number and the fact, that no
information on cascade production is available, prevents us from
separating the strange and baryon chemical potentials and to proceed
quantitatively with the analysis. Still, it is clear that $\mu_{\rm B}$
should be smaller than in S--W collisions, reducing the sensitivity of
our method which works better for baryon-rich environments. On the other
hand, forthcoming results from recent measurements in the S--Pb collision
system by the NA36 collaboration \cite{NA36}, which cover a larger
rapidity window, promise to provide the possibility of verification of
the picture developed here. 
 
The above discussion of the entropy content of the fireball and its
relation to the particle ratios demonstrates that the existing S--W data
are not only inconsistent with a fireball consisting of a HG with ${\cal
S/B}\sim 20$, but suggest strongly a QGP interpretation of the high
entropy phase: we saw that the EMU05 data \cite{EMU05} imply a source
with specific entropy of order 50, showing a remarkable coincidence with
the value computed for a QGP state with the thermal parameters determined
for the WA85 strange particle multiplicities \cite{WA85}. The necessity
of a high entropy phase as the source of the abundant strange particles
observed by WA85 \cite{WA85}, NA35 \cite{NA35}, and NA36 \cite{NA36}
provides a strong indication for the formation of a hot baryon-rich
quark-gluon plasma in 200 GeV\,A S$\to$A collisions. If true this should
provide a very interesting physics environment for the upcoming
experiments with Au and Pb beams. 
 
%%%%%%%%%%%%%%%%%%%%%%%%%%%%%%%%%%%%%%%%%%%%%%%%%%%%%%%%%%%%%%%%%%%%%


\begin{thebibliography}{99}
\bibitem{fn1}
In Ref. \cite{LTHSR} we have recently shown that these conclusions of 
Ref. \cite{Raf91} are insensitive to the production and decay of 
resonances and to the possible presence of a flow component in the 
transverse momentum spectra.
\end{thebibliography}
\end{document}